\documentclass[onecolumn,tightenlines,amsmath,amssymb,11pt,superscriptaddress,nofootinbib]{revtex4}

\usepackage{graphicx}
\usepackage{amsmath,amssymb,amsfonts,amsthm,stmaryrd,mathtools,bm,physics}
\usepackage{soul}
\usepackage{color}
\usepackage{tikz}
\allowdisplaybreaks[1]
\usepackage[bookmarks,linktocpage, colorlinks=true, plainpages = false, citecolor = treegreen,  linkcolor=darkblue, urlcolor = darkblue, filecolor = blue]{hyperref} 

\definecolor{tealgreen}{rgb}{0.0, 0.5, 1.0}
\definecolor{darkblue}{rgb}{0., 0.4, 0.8}
\definecolor{cadmiumred}{rgb}{1., 0., 0.22}
\definecolor{treegreen}{rgb}{0., 0.7, 0.3}

\usepackage{pifont}

\def\be#1\ee{\begin{align}#1\end{align}}

\def\ba{\begin{eqnarray}}
\def\ea{\end{eqnarray}}
\def\nn{\nonumber}

\begin{document}

\title{A non-local way around the no-global-symmetries conjecture in quantum gravity?
}

\author{Johanna Borissova}
\email[Electronic address (corresponding author): ]{jborissova@pitp.ca}
\affiliation{Perimeter Institute for Theoretical Physics, 31 Caroline Street North, Waterloo, ON, N2L 2Y5, Canada}
\affiliation{Department of Physics and Astronomy, University of Waterloo, 200 University Avenue West, Waterloo, ON, N2L 3G1, Canada}
\author{Astrid Eichhorn}
\email{eichhorn@sdu.dk} 
\affiliation{CP3-Origins, University of Southern Denmark, Campusvej 55, 5230 Odense M, Denmark}
\author{Shouryya Ray}
\email{shouryyar@setur.fo} 
\affiliation{CP3-Origins, University of Southern Denmark, Campusvej 55, 5230 Odense M, Denmark}
\affiliation{Department of Science and Technology, University of the Faroe Islands, Vestara Bryggja 15,
FO-100 T\'{o}rshavn, Faroe Islands}

\begin{abstract}
The no-global-symmetries conjecture is central to the swampland program that delineates the boundary between effective field theories that can be obtained from a quantum theory of gravity to those that cannot.  The conjecture states that virtual black-hole configurations in the path integral generate terms that violate  all global symmetries in the effective action for matter. Because of its central role, it is crucial to understand limitations to the validity of this conjecture.
In the context of the Lorentzian path integral over spacetime geometries, we explore whether virtual black-hole configurations can be suppressed dynamically. 
To that end, we work in a spherically symmetric setting and make use of horizon-detecting curvature invariants which vanish on the horizon. By constructing a non-local gravitational action from the inverse of such curvature invariants, we can achieve destructive interference of black-hole configurations in the path integral. Given that non-local gravitational actions appear generically as the result of integrating out matter degrees of freedom from a theory for quantum gravity and matter, our exemplary construction reinforces discussions about the role of non-locality in assessing arguably universal properties of quantum gravity within the framework of path integrals.
\end{abstract}

\maketitle
\tableofcontents

\section{Introduction: Black-hole configurations in the Lorentzian path integral}\label{Sec:Introduction}
 
In this note, we consider
black-hole configurations in the gravitational path integral and investigate how to suppress them dynamically. Our motivation stems from their central role in the no-global-symmetries conjecture~\cite{Banks:1988yz, Giddings:1987cg, Lee:1988ge, Abbott:1989jw, Coleman:1989zu, Kamionkowski:1992mf, Holman:1992us, Kallosh:1995hi, Banks:2010zn}, which is one of the swampland conjectures~\cite{Vafa:2005ui, Ooguri:2006in}, see~\cite{Brennan:2017rbf,Palti:2019pca,vanBeest:2021lhn,Grana:2021zvf,Agmon:2022thq} for reviews. It is often viewed as one of the best-founded conjectures, as it is based purely on semi-classical arguments.  Put simply, the conjecture states that if a black hole evaporates completely by emitting Hawking radiation, with its entropy described by the Bekenstein-Hawking area law, and does not leave behind a remnant, then continuous global symmetries cannot be conserved: Black holes can form by gravitational collapse from matter transforming in an arbitrarily complicated representation of a global symmetry group. Therefore, if they evaporate completely, the global charges are no longer present in the universe. If one thus considers an initial spacelike hypersurface $\Sigma_1(t_1)$ at a time $t_1$ before a black hole has formed and a final spacelike hypersurface $\Sigma_2(t_2)$ at a time $t_2$ after its complete evaporation, then the divergence of the current encoding the global symmetry is non-zero. 

At the quantum level, the contribution of virtual black holes to the path integral thus generates effective interactions for the matter theory which do not obey any global symmetries. In the Standard Model, this has consequences for the lifetime of the proton, whereas beyond the Standard Model, consequences, e.g., for axion-physics are well-explored~\cite{Alvey:2020nyh}, see the review~\cite{Reece:2023czb} and some phenomenologically popular models of dark matter are also in fact incompatible with the conjecture, see, e.g.,~\cite{Shiu:2013wxa}, see~\cite{Draper:2022pvk} for a recent review.~\footnote{We caution that in parts of the literature, no distinction is being made between the relative swampland of a given quantum-gravity theory (e.g., string theory) and the absolute swampland pertaining to all theories of quantum gravity, see~\cite{Eichhorn:2024rkc} for a discussion of the distinction.}
In the context of the AdS/CFT correspondence, the no-global-symmetries conjecture, which as stated above applies to continuous symmetries, has been extended to discrete symmetries~\cite{Harlow:2018tng}. In string theory, the no-global-symmetries conjecture holds for continuous symmetries because, put simply, every global symmetry automatically comes together with a massless vector field that turns out to be the gauge field of the corresponding local symmetry~\cite{Banks:1988yz}. In contrast, in asymptotic safety, numerous explicit calculations based on the Euclidean path integral show that global symmetries that are postulated for the matter sector are generically left intact by quantum-gravity fluctuations, e.g., \cite{Eichhorn:2011pc,Labus:2015ska,Meibohm:2016mkp, Eichhorn:2016vvy, Eichhorn:2017eht,Daas:2021abx,Eichhorn:2020sbo,deBrito:2021pyi, Laporte:2021kyp, deBrito:2023kow}, 
with consequences, e.g., for the lifetime of the proton~\cite{Eichhorn:2023jyr},
see~\cite{Eichhorn:2022gku} for a review. To be maximally conservative about the interpretation of these results (which are subject to systematic uncertainties), we conclude that there is therefore currently no evidence \emph{for} the no-global-symmetries conjecture in asymptotically safe gravity. Furthermore, in simplified, typically lower-dimensional settings, violations of the no-global-symmetries conjecture are known, and reviewed, e.g., in~\cite{Harlow:2020bee}, but these are not candidate theories of quantum gravity that pertain to our universe.
In other four-dimensional candidate quantum-gravity theories, the status of the no-global-symmetries conjecture is unclear and it is thus unknown whether this conjecture defines an absolute swampland that applies to all quantum-gravity approaches, see~\cite{Eichhorn:2024rkc} for a discussion. 

Ultimately, it is up to experiments to determine (i) whether global symmetries exist in nature and (ii) if they are violated, how large the violation is, see, e.g., \cite{Alvey:2021hjp}. However, it is expected that violations are suppressed by a dimensionless factor $e^{-M_{\rm Planck}^2/\Lambda^2}$~\cite{Fichet:2019ugl,Daus:2020vtf}, where $\Lambda$ is the cutoff of the effective field theory, in addition to a suppression by $E/M_{\rm Planck}$ to some power for interactions that are irrelevant. Therefore, even upcoming experimental tests of the lifetime of the proton will not yet reach the Planck scale~\cite{Hyper-Kamiokande:2018ofw,DUNE:2016hlj,JUNO:2015zny} and for the time-being we therefore only have theoretical considerations to guide us.

To provide further insight into the no-global-symmetries conjecture, we explore whether there is a  way to remove black-hole configurations from the path integral by dynamical suppression. 
This brings us to the broader question of the Lorentzian path integral in quantum gravity.

From a general perspective, it is a central challenge of quantum-gravity research to understand the Lorentzian path integral.
The action which enters the phase factor $e^{i S}$  is approximated by the Einstein-Hilbert action $S_{\rm EH}$ at the semi-classical level. In symmetry-reduced settings, saddle-point approximations combined with techniques of contour deformation enable explicit calculations of transition amplitudes~\cite{Feldbrugge:2017kzv,Feldbrugge:2017mbc,DiazDorronsoro:2018wro,Feldbrugge:2018gin,Janssen:2019sex,Lehners:2023yrj, Dittrich:2021gww,Asante:2021phx,Borissova:2023izx,Borissova:2024txs}.  
Beyond the semi-classical level, however, the path integral based on $S_{\rm EH}$ breaks down due to the perturbative non-renormalizability of quantum Einstein gravity~\cite{tHooft:1974toh,Goroff:1985th,vandeVen:1991gw}. Unless one gives up on the path integral completely, several questions arise. First, what is the proper gravitational configuration space to sum over? Different quantum-gravity approaches provide different tentative answers to this question, ranging from the space of all geometries at fixed topology, see, e.g., \cite{Ambjorn:2024pyv}, to the space of all geometries on all topologies~\cite{Gurau:2024nzv}, to the sum over all possible discrete causal structures for any dimension and any topology~\cite{Surya:2019ndm}, although a rigorous understanding of most of these configuration spaces is to date lacking.~\footnote{Even for fixed topology and dimensionality, many questions on gravitational configuration spaces remain open~\cite{Knorr:2022mvn}.} The second question which arises concerns the choice of dynamics. Here again, the answers of distinct quantum-gravity approaches differ. These two questions are unquestionably intertwined, because a given configuration can be suppressed either by a different choice of measure $\mathcal{D}g_{\mu\nu}$, or by a different choice of action appearing in $e^{i S}$.

Against this background, we take a different approach to the gravitational path integral. We remain agnostic as to what constitutes the particular quantum-gravity candidate theory. Instead, we explore under which conditions we can suppress black-hole configurations in the path integral dynamically in order to circumvent the no-global-symmetries conjecture. In~\cite{Borissova:2020knn,Borissova:2023kzq}, some of us already observed that singular black-hole configurations can be suppressed dynamically, see also~\cite{Giacchini:2021pmr, Chojnacki:2021ves}. The main idea is that if the action diverges when evaluated on a subset of ``neighboring" configurations~\footnote{Here, by a neighboring configuration we refer to one that can be reached by an infinitesimal change in one of the metric functions as expanded upon further in Section~\ref{Sec:EffectiveActionNonlocality}. We leave aside questions regarding the topology and local geometry of the space of four-dimensional spacetimes at a diffeomorphism-invariant level.}, the phase factor $e^{iS}$ oscillates infinitely fast. Thus, the configurations interfere destructively and do not contribute to the path integral.
A related idea was previously explored for cosmology in~\cite{Lehners:2019ibe,Lehners:2023fud}. Based on the finite-action principle~\cite{Barrow:1988gzc,Barrow:2019gzc}, proposed as a selection principle for physically viable classical solutions, 
it was shown that the requirement of a finite 
quadratic-curvature action disfavours anisotropic and inhomogeneous early universes~\cite{Lehners:2019ibe,Lehners:2023fud}.

Since the curvature singularity of classical black holes is usually considered irrelevant for the no-global-symmetries conjecture, a suppression that selectively targets singular black-hole configurations, as explored in~\cite{Borissova:2020knn,Borissova:2023kzq}, is not sufficient, even though specific higher-order derivative terms in the dynamics can even suppress certain regular black holes~\cite{Giacchini:2021pmr,Borissova:2023kzq}.
Instead, 
we target the part of a black-hole spacetime that is at the core of the no-global-symmetries conjecture: the horizon of a black hole. We use scalar curvature invariants \cite{Shakerin:2021dph} which detect horizons and investigate how to incorporate them in a dynamics that suppresses horizons in the path integral via destructive interference.

An apparent horizon defines a quasi-local surface which can be detected locally by means of scalar polynomial invariants of the Riemann curvature tensor which change sign on the horizon surface~\cite{Karlhede:1982fj,Tammelo:1997sh,Gass:1998nd,Mukherjee:2002ba,Lake:2003qe,Saa:2007ub,Gomez-Lobo:2012ibv,Moffat:2014aqa,Visser:2014zqa,Abdelqader:2014vaa,Page:2015aia,McNutt:2017gjg,McNutt:2017paq,Coley:2017vxb,McNutt:2021esy}. Such horizon-detecting invariants in turn allow us to construct gravitational actions which are highly oscillatory for configurations potentially violating global symmetries due to the existence of a black-hole horizon. 

 This note is structured as follows. First, in Section~\ref{Sec:Invariants} we review  
the invariant characterization of spherically symmetric black-hole horizons via scalar polynomial curvature invariants. Subsequently, in Section~\ref{Sec:EffectiveActionNonlocality} we construct non-local gravitational actions which 
produce highly oscillatory phase factors for spherically symmetric black-hole configurations and discuss how to localize the action. Finally, we finish with a discussion and outlook in Section~\ref{Sec:Discussion}.

\section{Invariant characterization of black-hole horizons}\label{Sec:Invariants}
 
In a first approach to black holes, one often focuses on the event horizon as their key characteristic property.  
The event horizon is, however, a global property of the full spacetime and its definition teleological as it requires knowledge of the entire future dynamical evolution of null geodesics~\cite{Booth:2005qc}. 
Apparent horizons~\cite{Ashtekar:2004cn,Booth:2005qc,Gourgoulhon:2008pu}, on the other hand, are quasi-local surfaces whose characterization
only requires local information about the expansion of outward and inward pointing null geodesics. For stationary black holes, event horizons are also apparent horizons~\cite{Booth:2005qc}. However, for dynamical black holes, these two types of horizons generically differ. The foliation-dependence and therefore non-uniqueness of quasi-local surfaces in time-dependent settings has motivated the search for invariant hypersurfaces independent of the choice of foliation. In this context the notion of geometric horizons~\cite{Coley:2017woz,Coley:2017vxb} has been introduced. A geometric horizon describes a surface on which the curvature tensor and its covariant derivatives become algebraically special and which can be defined invariantly through the vanishing of a certain set of curvature invariants~\cite{Coley:2017woz,Coley:2017vxb}. For spherically symmetric black-hole spacetimes, the apparent horizon is also a geometric horizon and in particular is unique~\cite{Coley:2017vxb}. 

The quasi-local definition of a geometric horizon in terms of curvature invariants also makes it possible to come up with an experimental prescription which would allow an observer to determine whether they are crossing a geometric horizon. The curvature invariants that define a geometric horizon can thus be understood as ``horizon detectors". To describe how such a horizon detector works, let us start from 
the curvature tensor, which measures the relative acceleration of test particles. In a Schwarzschild black-hole spacetime, no qualitative changes of the geometry occur in the vicinity of an observer who passes through the horizon at $r_H=2M$. Monitoring the
value of the Kretschmann scalar, given by the square of the Riemann curvature tensor,
\be
R_{\mu\nu\rho\sigma}R^{\mu\nu\rho\sigma} = \frac{48 M^2}{r^6}
\ee
provides information about the local curvature in the Schwarzschild spacetime but does not allow the observer to detect the passage through the apparent horizon, unless they have a priori knowledge about the global mass parameter $M$. By contrast, there exist other local invariants constructed from the curvature tensor and its covariant derivatives, which can probe a neighborhood of a stationary surface and provide horizon-detectors in large classes of black-hole spacetimes~\cite{Karlhede:1982fj,Tammelo:1997sh,Gass:1998nd,Mukherjee:2002ba,Lake:2003qe,Saa:2007ub, Gomez-Lobo:2012ibv,Moffat:2014aqa,Visser:2014zqa,Abdelqader:2014vaa,Page:2015aia,McNutt:2017gjg,McNutt:2017paq,Coley:2017vxb,McNutt:2021esy}. The horizon-detecting feature of certain scalar invariants is reflected in their sign change on the black-hole apparent horizon. For example, the square of the gradient of the Riemann curvature tensor can provide such an invariant. When evaluated for the Schwarzschild geometry, this invariant is given by
\be
\nabla_\alpha R_{\mu\nu\rho\sigma}\nabla^\alpha R^{\mu\nu\rho\sigma} = \frac{720 M^2}{r^8} \qty(1-\frac{2 M}{r})\,.
\ee 
It is smooth for all $r\in (0,\infty)$, positive in the black-hole exterior $r>2M$, negative in the interior $r<2M$, and vanishes only on the horizon surface where $r_H=2M$. Such a property implies that an infalling observer can in principle detect the presence of an apparent horizon in the Schwarzschild spacetime quasi-locally in a finite-size laboratory. Using the relativistic quadratic geodesic deviation equation, the above invariant can be expressed in terms of measurable quantities associated with the separation, relative velocity and acceleration of test particles in the Schwarzschild spacetime~\cite{Tammelo:1997sh}. 

In what follows we demonstrate the possibility of physical detection of the apparent horizon of generic spherically symmetric black holes. To that end, we consider the general
spherically symmetric non-stationary line element  in advanced coordinates 
given
by~\cite{Senovilla:2011fk}
\be\label{eq:MetricImplodingSphericalSymmetry}
\dd{s^2}_{\rm sph.~symm.} = -e^{{2}\beta(v,r)}\qty(1-\frac{2m(v,r)}{R(r)})\dd{v}^2 + 2 e^{\beta(v,r)} \dd{v}\dd{r} + R(r)^2   \dd{\Omega}^2\,,
\ee
where $\dd{\Omega}^2 =  \dd{\theta}^2+ \sin^2(\theta) \dd{\phi}^2$ is the area element on the unit two-sphere $\mathcal{S}^2$. The special case of a stationary black-hole spacetime is given by the choice $\beta(v,r)= \beta(r)$ and $m(v,r)=m(r)$. Generically, the a priori arbitrary mass function $m(v,r)$ and functions $\beta(v,r)$ and $R(r)$ can be chosen such that the line element~\eqref{eq:MetricImplodingSphericalSymmetry} describes a dynamical spherically symmetric black hole or wormhole~\cite{McNutt:2021esy}. In both cases the black-hole apparent horizon or wormhole throat, respectively, are described by the equation $R(r_H) = 2 m(v_H,r_H)$ which follows from $g_{vv}=0$.

The existence of a black-hole apparent horizon affects the structure of the curvature tensor and its covariant derivatives. As for vacuum black holes in GR the Ricci tensor vanishes everywhere, it is evident that these invariants must involve the Riemann or Weyl curvature tensor. Most invariants discussed in the literature use the Weyl tensor, which is conformally invariant. Given that an apparent horizon is a property of the causal structure and cannot be removed by a conformal transformation, it can be expected that information on the Weyl curvature tensor and its derivatives is sufficient to detect apparent horizons.
Specifically, the invariant 
\be\label{eq:Chi}
\chi=4 C^2 \qty(\nabla_\mu C)^2 - \qty(\nabla_\mu C^2)^2\,,
\ee
 where $C_{\mu\nu\rho\sigma}$ denotes the Weyl tensor,
is a horizon detector for spherically symmetric black holes (and wormholes) described by the line element Eq.~\eqref{eq:MetricImplodingSphericalSymmetry}~\cite{Coley:2017vxb,McNutt:2021esy}. \footnote{ This statement follows from the observation that the value of this invariant is proportional to the product of the future null expansions relative to an appropriately chosen null frame, and this product vanishes on the apparent horizons of the line element~Eq.~\eqref{eq:MetricImplodingSphericalSymmetry}~\cite{McNutt:2021esy}.} In other words, this invariant vanishes on the surface $R(r) = 2 m(v,r)$, but is generically non-zero away from it, as can be seen by an explicit evaluation of $\chi$ on the class of spacetimes of the form Eq.~\eqref{eq:MetricImplodingSphericalSymmetry},
\be\label{eq:ChiEvaluated}
\eval{\chi}
_{\dd{s^2}_{\rm sph.~symm.}
} = \qty(R(r) - 2 m(v,r)) \mathcal{F}(m(v,r),\beta(v,r),R(r))\,,
\ee
where
\ba
\mathcal{F} &=& \frac{128 R'^2 }{3 R^{15}}e^{-4\beta} \Bigl(-6 e^\beta m R'^2 + e^\beta R \qty(-1+ R'^2 + 3 m R'' + R' \qty(4 m' + 5 m  \beta')) \nonumber\\
&{}&- e^\beta R^2 \qty(R'' + R' \beta' + 3 m' \beta' + 2 m \beta'^2 +m'' + 2 m \beta'') + R^3 \qty(e^\beta \beta'^2 + e^\beta \beta'' +  \dot{\beta}')\Bigr)^4\,.
\ea

The invariant $\chi$ provides the main ingredient for our analyses on the dynamical suppression of spherically symmetric black-hole configurations in the path integral in Section~\ref{Sec:EffectiveActionNonlocality}.

\section{Gravitational actions and non-locality}\label{Sec:EffectiveActionNonlocality}

In the following we consider gravitational actions for a quantum-gravity path integral which dynamically suppresses global-symmetry-violating configurations. We will focus specifically on the suppression of spherically symmetric black holes in the path integral. To that end, our aim is to construct a gravitational action which diverges on neighboring black-hole configurations of this type, such that the infinitely rapid change of the phase factor from one configuration to a neighboring one leads to destructive interference between these. An analogous construction for the dynamical suppression of spacetimes with a curvature singularity has been studied in~\cite{Borissova:2020knn,Borissova:2023kzq}. 

This section is subdivided into three subsections. First, we introduce a scalar invariant as the inverse of a horizon detector for spherically symmetric black-hole spacetimes and illustrate that it allows for dynamical suppression of neighboring back-hole configurations in a way specified further in Section~\ref{SecSub:DestructiveInterference}. Second, in Section~\ref{SecSub:ContributionFLRWMinkowski} we discuss a modification of the invariant which allows physical non-black-hole spacetimes such as homogenous and isotropic FLRW spacetimes and the flat Minkowski spacetime, to contribute to the path integral. Finally, we consider a localization of the effective gravitational actions by means of auxiliary matter fields in Section~\ref{SecSub:Localization}. 

\subsection{Destructive interference of spherically symmetric black holes via horizon-detecting invariants}\label{SecSub:DestructiveInterference}

To investigate the dynamical suppression of spherically symmetric black holes in the path integral, we will make use of the invariant defined in equation~\eqref{eq:Chi}, i.e.,
\ba\label{eq:ChiAgain}
\chi \equiv  4 C^2 \qty(\nabla_\mu C)^2 - \qty(\nabla_\mu C^2)^2 \,,
\ea
 where we remind the reader that $C_{\mu\nu\rho\sigma}$ denotes the Weyl tensor.
We have shown in Section~\ref{Sec:Invariants} and Eq.~\eqref{eq:ChiEvaluated}, that $1/\chi$ diverges on the horizon of any spherically symmetric black-hole spacetimes described by the line element~\eqref{eq:MetricImplodingSphericalSymmetry}, whose horizons are defined by solutions to $R(r)-2m(v,r)=0$. Therefore $1/\chi$ provides a suitable invariant for a gravitational action which dynamically suppresses spherically symmetric black-hole spacetimes in the path integral via destructive interference.

For destructive interference to take place, it is central that the action does not diverge only on a single point in configuration space, but on entire families of neighboring black-hole configurations. Verifying that this is the case would require a precise notion of ``black-hole neighbors" in configuration space. As gravitational configuration spaces are generically only poorly understood, we proceed more heuristically to define neighboring black-hole configurations. We use
that $\chi$ becomes zero on the surface $R(r)-m(v,r)=0$ for the spherically symmetric spacetimes~\eqref{eq:MetricImplodingSphericalSymmetry}, independently of the specific choices of mass function $m(v,r)$ and functions $\beta(v,r)$ and $R(r)$, cf.~Eq.~\eqref{eq:ChiEvaluated}. 

To that end, let us consider an initial given black-hole spacetime with fixed functions $R(r)$, $m(v,r)$ and $\beta(v,r)$, for which $1/\chi$ diverges on the horizon surface $R(r)-m(v,r)=0$. Then we may infinitesimally deform the black-hole spacetime into another black-hole spacetime by modifying the functions $R$, $m$ and $\beta$ into new functions $\tilde{R}$, $\tilde{m}$ and $\tilde{\beta}$, such that the equation $\tilde{R}(r)-\tilde{m}(v,r)=0$ still has a solution. 
The latter guarantees that the deformed spacetime still describes a black-hole spacetime.~\footnote{In principle, the modified spacetime may also be a wormhole spacetime. Since we are only interested in infinitesimal deformations, we exclude this possibility.} Modifications of the functions $R(r)$ and $m(v,r)$ can be understood as deformations of the horizon surface of the black-hole spacetime, whereas modifications of the function $\beta(v,r)$ describe a deformation of the spacetime away from the horizon. This allows us to view the deformed spacetime as a ``neighbor" of the initial black-hole spacetime. Moreover, from the explicit form of $\chi$ evaluated on the general line element~\eqref{eq:MetricImplodingSphericalSymmetry}, it is clear that $1/\chi$ will also diverge on the deformed black-hole configuration. At the level of the path integral built from an action involving $1/\chi$, we therefore expect destructive interference between these and other neighboring black-hole configurations to occur.

\begin{figure}[t]
	\centering
	\includegraphics[width=.51\textwidth]{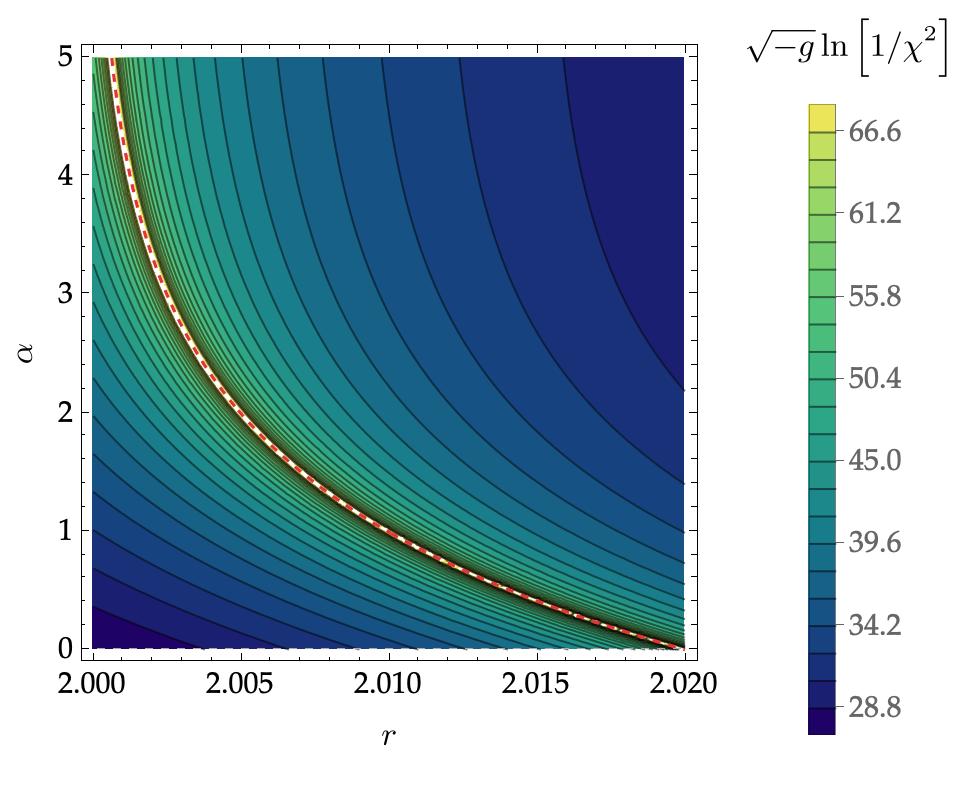}
	\hfill
		\includegraphics[width=.47\textwidth]{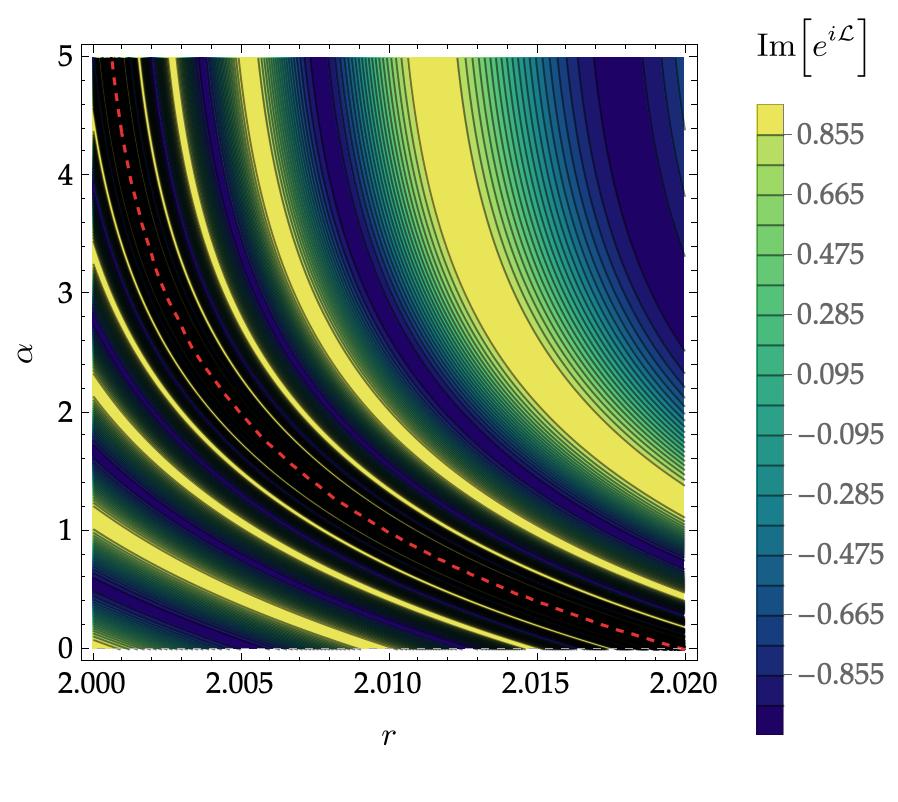}\\
\includegraphics[width=0.47\textwidth]{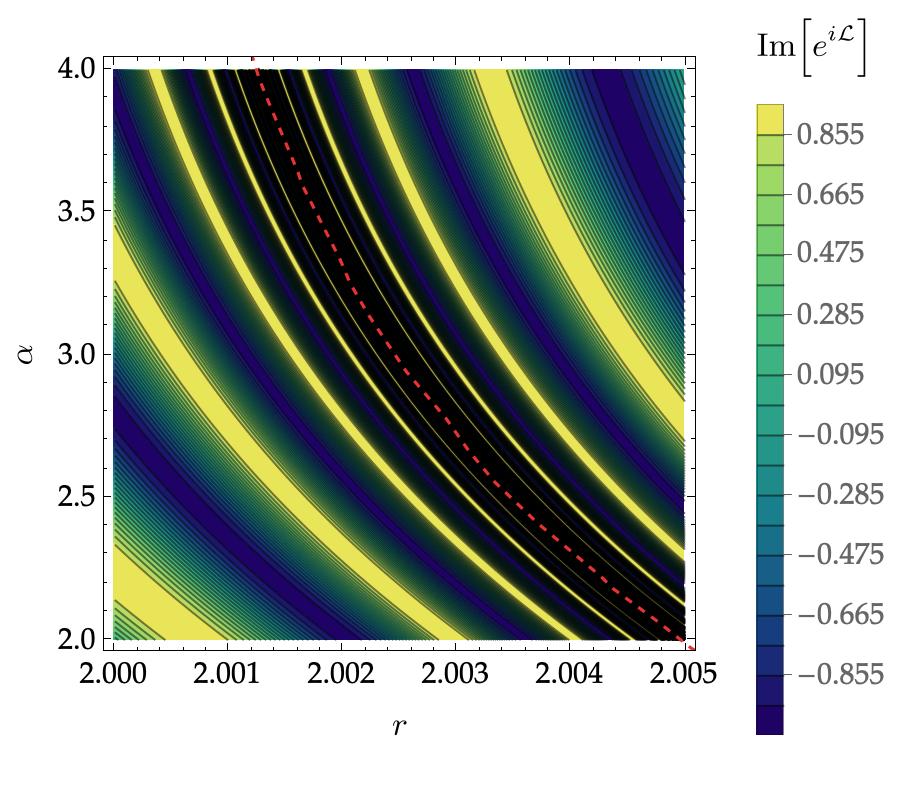}\quad\quad\includegraphics[width=0.47\textwidth]{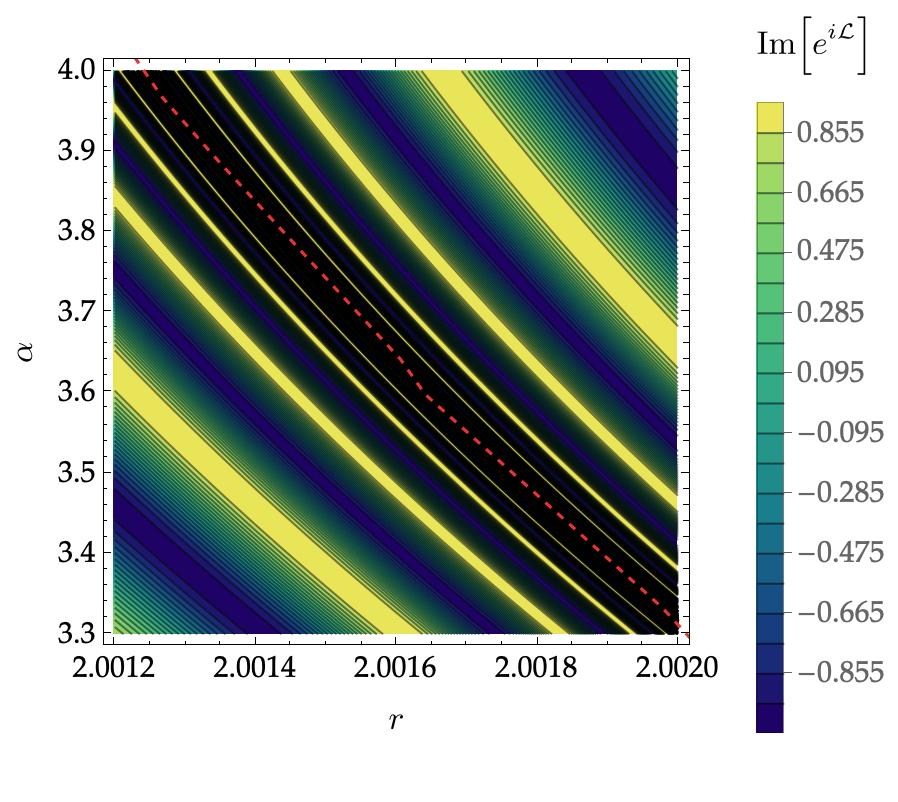}
	\caption{\label{Fig:LagrangianDeformedSpacetimes} Upper Left: We show the Lagrangian density~\eqref{eq:LExample} for black-hole configurations obtained by infinitesimal one-parameter deformations of the Schwarzschild spacetime via the replacement $M \mapsto M + m_\alpha r^\alpha$, for specific values $M\equiv 1$ and $m_\alpha \equiv 1/100$, as a function of the radial coordinate $r$ and effective deformation parameter $\alpha$ in the $\theta = \pi/2$ plane (as explained in the main text). Along the red dashed contour, $(r-2M)r^\alpha -2 m_\alpha =0$ holds, i.e., this contour marks the horizons of the deformed black-hole spacetimes. The Lagrangian density becomes highly divergent on the horizons of all such neighboring black-hole configurations. Upper Right: We show the imaginary part of the path-integrand factors $e^{iS}$, for simplicity without explicitly performing the spacetime integral in the action, i.e., $\text{Im}\qty[e^{i\mathcal{L}}]$. It is highly oscillating for all neighboring black-hole configurations and thereby expected to result in the destructive interference between these. Lower panels: We show zooms into the region plotted in the right upper panel to highlight that the imaginary part of $e^{i \mathcal{L}}$ oscillates infinitely fast close to the horizon, i.e., the plotted structure approaches self-similarity.}
\end{figure}

As a concrete example, we can consider a subclass of static spherically symmetric black holes  described
by the line element~\eqref{eq:MetricImplodingSphericalSymmetry} with the choice of functions $e^{\beta(r,v)} = 1$, $R(r) = r$ and $m(v,r)=m(r)$, i.e.,
\be\label{eq:MetricExample}
\dd{s^2} = - g_{vv} \dd{v}^2 + 2 g_{vr} \dd{v}\dd{r} + r^2   \dd{\Omega}^2= -\qty(1-\frac{2m(r)}{r})\dd{v}^2 + 2 \dd{v}\dd{r} + r^2   \dd{\Omega}^2\,.
\ee
Here the mass function $m(r)$ is an arbitrary function of the radial coordinate $r$. The apparent horizon(s) of a black hole described by the line element~\eqref{eq:MetricExample}, are defined by the equation $g_{vv}=0$, i.e., $r_H-2m(r_H)=0$. An explicit evaluation of the invariant~$\chi$ in~\eqref{eq:ChiAgain} gives
\be
\chi =  \qty(r-2m(r))\qty(6 m(r)+r\qty(-4 m'(r) + r  m''(r)))^4\cdot \frac{128}{3 r^{15}}\,.
\ee
Starting from a fixed configuration describing a Schwarzschild black hole with mass $m(r)=M$ and horizon located at $r_H=2M$, we may deform the spacetime into other asymptotically flat black-hole spacetimes~\footnote{Asymptotic flatness of a black-hole spacetime in this example refers to the mass function satisfying $m(r)\to 0$ for $r\to \infty$. We note that an action built from $1/\chi$ will generically diverge at infinity for asymptotically flat spacetimes. Such an infrared (IR) divergence can be regularized
by restricting the integration range for the radial coordinate by means of an IR cutoff and is not relevant to our discussion. Later in Section~\ref{SecSub:ContributionFLRWMinkowski} we will consider a refined version of an inverse horizon-detecting invariant, for which the action remains well-behaved for large classes of asymptotically flat spacetimes.}, by adding, e.g., negative powers of $r$ to the mass function,
\be
m(r) = M + \sum_{\alpha >0} M_\alpha r^{-\alpha}\,
\ee
with real coefficients $M_\alpha$. If we assume only one of these $M_\alpha$ to be non-vanishing, for different $\alpha$, we obtain infinitesimal one-parameter deformations of the Schwarzschild spacetime by the prescription $M_\alpha = 0 \mapsto M_\alpha = m_\alpha$ with $m_\alpha \ll 1$. The horizons of these deformed black-hole spacetimes labeled by parameters $\alpha$ satisfy
\be
\qty(r_{H_\alpha} - 2M) \qty(r_{H_\alpha})^\alpha - 2 m_\alpha = 0\,.
\ee
They are detected by the invariant $\chi$ which, evaluated on the deformed spacetimes explicitly, takes the form
\be
\chi_\alpha = \qty((r - 2M) r^\alpha - 2 m_\alpha) \qty(6 M r^\alpha + m_\alpha \qty(\alpha^2 + 5\alpha + 6))^4  \frac{128}{3 r^{5(\alpha+3)}}\,.
\ee
The path-integral configuration space restricted to the set of variables $\{m_\alpha\}$ is infinite-dimensional. To provide a pictorial representation of the described infinitesimal deformations around a given Schwarzschild black-hole configuration, we fix exemplary $M\equiv 1$ and $m_\alpha \equiv 1/100$, whereby the smallness of the last value represents the infinitesimality of the deformation with respect to the gravitational configuration space $({m_\alpha})$. Consequently, the deformations can be effectively described by the continuous parameter $\alpha$. Herewith, in the left plot of Fig.~\ref{Fig:LagrangianDeformedSpacetimes}, we depict a potential horizon-detecting Lagrangian density
\be \label{eq:LExample}
\mathcal{L} \equiv \sqrt{-g}\ln(1/\chi^2),
\ee
whose explicit form will be motivated further in Section~\ref{SecSub:Localization}, as function of the radial coordinate $r$ and deformation parameter $\alpha$ in the $\theta = \pi/2$ plane. The Lagrangian density $\mathcal{L}_\alpha$ becomes highly divergent along the red dashed contour defined by $(r-2M)r^\alpha -2 m_\alpha =0$, i.e., on the horizons of all of the deformed black-hole spacetimes. Accordingly, the path-integrand factors $e^{i S}$ (without the spacetime integral carried out) as shown in the right plot of Fig.~\ref{Fig:LagrangianDeformedSpacetimes}, oscillate rapidly on all of these neighboring black-hole configurations and are expected to result in destructive interference between these in the overall gravitational path integral.

\subsection{Contribution of physical spacetimes}\label{SecSub:ContributionFLRWMinkowski}

We cannot simply use $1/\chi$ as building block for the action, since $\chi$ evaluates to zero, i.e., $1/\chi$ diverges for any Weyl-flat spacetime and in particular for Weyl-flat spacetimes which are not black-hole spacetimes. We thus need to refine our proposal for an action.

On this basis, we will instead consider the following invariant as building block of a Lagrangian density,
\be\label{eq:JInv}
\mathcal{J} \equiv \frac{\qty(C^2)^{4 }}{\chi}\,.
\ee
In a suitable limit, this invariant vanishes on Weyl-flat spacetimes, since the total power of Weyl tensors in the numerator is higher than in the denominator. As a specific example, we can consider Minkowski spacetime and FLRW spacetimes, which we construct as the limit of a modified FLRW spacetime defined by
\be
\dd{s}^2= -\qty(1+ c\, r^2)dt^2 + a(t)^2(dr^2 + r^2 d\Omega^2)\,.
\ee
Here $c$ is an arbitrary non-vanishing dimensionful real parameter which
renders the spacetimes non-Weyl-flat. We get back to standard FLRW with scale factor $a(t)$ in the limit $c\rightarrow 0$ and obtain Minkowski spacetime in the limits $c \to 0$ and $a(t) \rightarrow \rm const$ $\neq 0$. On this class of spacetimes,
\ba
C^2 &=& \frac{4c^4r^4}{3a(t)^4 \qty(1+ c r^2 )^4}\,,\\
\chi &=& \frac{128 c^8 r^6 \qty(1 + cr^2 - r^2 a'(t)^2)}{2 a(t)^{10}\qty(1+cr^2)^9}\,.
\ea
We can thus evaluate $\mathcal{J}$ in equation~\eqref{eq:JInv} for FLRW by taking the limit $c \to 0$,
\be
\mathcal{J}_{\,\rm FLRW} = \lim_{c\to 0} \frac{2 c^8 r^{10}}{27 a(t)^6 \qty(1+c r^2)^7 \qty(1 + c r^2 - r^2 a'(t)^2)} = 0\,.
\ee
We have thereby constructed a contribution to the action which vanishes on cosmological FLRW spacetimes and flat spacetime, but diverges whenever $\chi=0$, but $C^2 \neq 0$, which is satisfied for general spherically symmetric black-hole horizons.

\subsection{Localization and effective actions}\label{SecSub:Localization}

An action built from the invariant $\mathcal{J}$ in Eq.~\eqref{eq:JInv} is non-local. While it is built from an overall positive number of derivatives, it contains derivatives in the denominator. Given the inherent non-local property of horizons, it is perhaps not surprising that a non-local action is needed to suppress black-hole spacetimes in the path integral, even if the horizon-detecting invariants themselves are local.

In what follows we consider a localization of the gravitational action built from the invariant $\mathcal{J}$. An action of the form $\pm i\, \int d^4x \sqrt{g} \ln \mathcal{I}$ can be localized with a scalar or Grassmann field. Thus, focusing on the invariant $\mathcal{J}$ in Eq.~\eqref{eq:JInv}, we transition to an action of the form
\be
S = \int d^4x \sqrt{g} \ln (\frac{\qty(C^2)^8}{\chi^2} ) \label{eq:Lchifinal}\,.
\ee
This action contains the logarithm of the square of $\mathcal{J}$. Our reason to consider the square of this invariant is that $\chi$ changes sign on a spherically symmetric black-hole horizon and would thereby make the action complex on one side of the horizon. 

Let us consider the following action for the metric field $g_{\mu\nu}$, an auxiliary real scalar field $\phi$, and an auxiliary pair of $d_\gamma$-component Grassmann fields $\psi,\bar{\psi}$:~\footnote{This construction is standard and generalizes from $\chi^2$ and $C^2$ to any other sufficiently well-behaved  
operator.}
\be \label{eq:ActionGravityAuxiliaryScalarField}
S[g_{\mu\nu},\phi,\psi,\bar{\psi}] = S_g[g_{\mu\nu}] + \frac{1}{2} \int \dd[4]x \sqrt{-g}\phi \chi^2 \phi + \int \dd[4]x \sqrt{-g}  \bar{\psi} C^2\psi\,.
\ee
Note that we make no assumptions about the representation of the Poincar\'e group under which $\psi,\bar\psi$ transform. These fields could be arranged, e.g., as Weyl or Dirac fermions. Similarly, the choice of internal structure is left unspecified. In the above action, $S_g$ represents the purely gravitational part of the action, and we have absorbed a free coupling in front of the interaction term via a rescaling of the field $\phi$, which in turn has mass dimension $[\phi] = 2 - \frac{[\chi^2]}{2}$. The absence of a kinetic term for $\phi$ implies that this field is non-dynamical. In other words, the redefined field $\Phi \equiv \phi^2$ acts as a Lagrange multiplier whose equations of motion impose the constraint $ \chi^2= 0$. Similar observations apply to $(\psi,\bar{\psi};C^2)$.\\

The Lorentzian path integral for the above gravity-matter system can be written as
\ba\label{eq:PathIntegralGravityAuxiliaryScalarField}
Z &=& \int \mathcal{D}g_{\mu\nu} \mathcal{D}\phi \mathcal{D}\psi \mathcal{D}\bar{\psi} \,e^{i S[g_{\mu\nu},\phi,\psi,\bar{\psi}]} = \int \mathcal{D}g_{\mu\nu} \, e^{i S_g[g_{\mu\nu}]}\qty(\int \mathcal{D}\phi\mathcal{D}\psi \mathcal{D}\bar{\psi} \, e^{i \int \dd[4]x \sqrt{-g} \left(\frac12 \phi \chi^2 \phi + \bar{\psi}C^2\psi\right)})\nn\\
&\equiv & \int \mathcal{D}g_{\mu\nu} \, e^{iS_\text{eff}[g_{\mu\nu}]}\,.
\ea
The last equality defines the effective gravitational action $S_\text{eff}[g_{\mu\nu}]$ which can be computed by evaluating the Lorentzian functional integral over the matter fields $\phi$, $\psi$ and $\bar{\psi}$. We focus in the following on the integral over $\phi$. To that end, writing the equations of motion for $\phi$ in operator form, $\hat{F}\phi = 0$ with $\hat{F}\equiv \chi^2$, the standard procedure is to consider the system in a finite box and impose periodic boundary conditions on the field $\phi$~\cite{Mukhanov:2007zz,Baldazzi:2019kim}. The spectrum of eigenvalues $\lambda_n $ of the eigenvalue problem $\hat{F}\phi_n = \lambda_n \phi_n$ is then discrete, the operator $\hat{F}$ self-adjoint and the eigenfunctions $\phi_n$ form a complete orthonormal basis with respect to the covariant scalar product $(\phi_n,\phi_m)\equiv \mu^{[\chi^2]}\int \dd[4]x \sqrt{-g} \phi_n(x)\phi_m(x) = \delta_{nm}$, where $\mu$ is a constant with dimension of mass. Expanding the field $\phi$ in the orthonormal basis $\{\phi_n\}$ of eigenfunctions of $\hat{F}$ in the form $\phi = \mu^{\frac{[\chi^2]}{2}-2}\sum_n c_n \phi_n$ with dimensionless coefficients $c_n$, the gravity-matter part of the action can be written as~\cite{Baldazzi:2019kim}
\be
\frac{1}{2}\int \dd[4]{x} \sqrt{-g}\phi \hat{F} \phi = \frac{1}{2 \mu^{[\chi^2]}}\sum_{n\in \sigma}  \lambda_n c_n^2\,.
\ee
Here $\sigma$ labels the set of eigenvalues of $\hat{F}$. The path-integral measure can be written formally as $\mathcal{D}\phi = \mathcal{N} \prod_{n\in \sigma} \dd c_n$ with an infinite field-independent normalization factor $\mathcal{N}$. The latter can be fixed, e.g., by imposing a Gaussian normalization condition on $\phi$~\cite{Baldazzi:2019kim}.
Consequently, the functional integral over the matter field $\phi$ becomes a product of Gaussian integrals with purely imaginary phase factors,
\be\label{eq:GaussianIntegralsProduct}
Z_\phi \equiv \int \mathcal{D}\phi \,e^{\frac{i}{2}\int \dd[4]{x} \sqrt{-g}\phi \hat{F} \phi} = \mathcal{N} \prod_{n\in \sigma} \int_\mathbb{R} \dd c_n \,e^{i\frac{\lambda_n}{2\mu^{[\chi^2]}} c_n^2}\,.
\ee 
For $\lambda_n > 0$ or $\lambda_n < 0$, the one-dimensional ordinary integrals over $c_n$ on the real axis are not well-defined. They can however be computed by deforming the integration contour into the complex plane and following the steepest-descent paths through the origin~\cite{Baldazzi:2019kim}. Similarly, for $\lambda_n =0$, $\chi^2$ has to be given a small imaginary part, $\chi^2 \to \chi^2+i \epsilon$, leading to $\lambda_n \to \lambda_n + i \epsilon$. As a consequence, the contribution of the zero eigenvalues reduces to a product of standard real Gaussian integrals. Altogether, the product of integrals in equation~\eqref{eq:GaussianIntegralsProduct} can be computed by decomposing the spectrum as $\sigma = \sigma_- \cup \sigma_0 \cup \sigma_+ $, where $\sigma_-$, $\sigma_0$, $\sigma_+$ correspond to negative, zero and positive eigenvalues, respectively, and subsequently evaluating each of the three types of integrals in the described way. With an appropriate normalization for the measure, the matter path integral~\eqref{eq:GaussianIntegralsProduct} can be expressed as a functional determinant~\cite{Baldazzi:2019kim},
\be
Z_\phi = \qty[\det(\frac{\hat{F} +i \epsilon}{\mu^{[\chi^2]}})]^{-\frac{1}{2}}\,.
\ee
Thus, in the absence of zero eigenvalues, one recovers the standard trace-log formula for the contribution of the matter path integral to the effective gravitational action~\cite{Baldazzi:2019kim}. The computation of the contribution from $(\psi,\bar{\psi})$ proceeds in analogous fashion, altogether leading to
\be\label{eq:Sefffinal}
S_{\text{eff}}[g_{\mu\nu}] = S_g[g_{\mu\nu}] + \frac{i}{2}\Tr \ln(\frac{\chi^2}{\mu^{[\chi^2]}}) - d_\gamma i \Tr \ln(\frac{C^2}{\mu^{[C^2]}}) = S_g[g_{\mu\nu}] + \frac{i}{2}\Tr \ln(\frac{\chi^2}{(C^2)^{2d_\gamma}})\,.
\ee

For $d_\gamma = 4$ we recover the action~\eqref{eq:Lchifinal}. The fact $d_\gamma = 4$ means, for instance, that the Grassmann contribution can be realised as arising from one Dirac fermion.\\

Let us close this section by pointing out potential problems which arise in localizing non-local actions involving horizon-detecting invariants, such as $\mathcal{J}$ in equation~\eqref{eq:JInv}, via auxiliary matter fields. The localization procedure necessarily makes use of the trace-log formula for the contribution of the matter path integral to the full quantum-gravity path integral~\eqref{eq:PathIntegralGravityAuxiliaryScalarField}. For this reason we considered the action~\eqref{eq:Lchifinal}, which contains the logarithm of the (squared) horizon-detecting invariant $\mathcal{J}$. The invariant $\mathcal{J}$ was constructed such that it is a horizon-detector for spherically symmetric spacetimes, but remains well-behaved and actually vanishes for Weyl-flat FLRW spacetimes, in a suitable limit described in Section~\ref{SecSub:ContributionFLRWMinkowski}. Taking the logarithm of $\mathcal{J}$ 
reintroduces a divergence of the action on these spacetimes, because the logarithm diverges not just at asymptotically large arguments, but also at zero. This divergence could be avoided by a trace-log contribution to~\eqref{eq:Sefffinal} of the form $\propto \Tr \ln(\alpha+\mathcal{J}^2 )$ where $\alpha$ is a positive real constant. Such an effective action would however not arise through a localization via non-dynamical matter fields. One may ask what the situation looks like if the auxiliary fields $\phi,\psi$ are promoted to dynamical degrees of freedom. 
 This amounts to supplying them with their respective kinetic terms, which on an operative level leads to the replacements $\chi^2 \to \square + \chi^2, (C^2)^2 \to \square + (C^2)^2$ in Eq.~\eqref{eq:Sefffinal}. %
A necessary condition for the construction to have the desired effect of suppressing horizons, is, if $0 \in \sigma(\square + \chi^2)$ is an accumulation point and $\mu_{\square + \chi^2}([0,r))$ grows fast enough with $r$ for small $r$, where $\sigma$ is the spectrum and $\mu$ is the spectral measure associated with $\square + \chi^2$. Meanwhile, the spectrum of $\square + (C^2)^2$ should not contain zero and its spectral measure should be sufficiently suppressed at large values. %
The spectrum of these two operators are, however, currently not well-understood, and we leave this for future work.

If such a construction goes through, then $\phi$ and $\psi$ may be thought of as matter fields and the coupling to $\chi^2$ and $C^2$ is a non-minimal interaction, as it would be expected in many (quantum) theories beyond General Relativity. This may suggest that a suppression mechanism as we consider here is much more prevalent than one might at first expect.

\section{Discussion}\label{Sec:Discussion}

In this note, we have brought together two lines of investigation: on the one hand, a body of previous literature is devoted to the construction of horizon-detecting curvature invariants for black holes~\cite{Karlhede:1982fj,Tammelo:1997sh,Gass:1998nd,Mukherjee:2002ba,Lake:2003qe,Saa:2007ub, Gomez-Lobo:2012ibv,Moffat:2014aqa,Visser:2014zqa,Abdelqader:2014vaa,Page:2015aia,McNutt:2017gjg,McNutt:2017paq,Coley:2017vxb,McNutt:2021esy}. On the other hand, previous studies explore which terms should be added to the dynamics such that black-hole spacetimes interfere destructively as a result of actions which diverge on spacetimes with curvature singularities~\cite{Borissova:2020knn,Borissova:2023kzq}. Here, we bring the reasoning on destructive interference due to a diverging action to bear on horizon-detecting curvature invariants. Since such invariants vanish on a black-hole horizon, the inverse horizon-detecting invariants must be used in the action. A generic action that achieves destructive interference is thus generically non-local. This is perhaps not a surprising property, given that a black-hole horizon is an inherently non-local concept.

We find, however, that it is necessary to make the non-locality milder, because the inverse of the  horizon-detecting invariant also diverges on simple cosmological spacetimes, such as homogeneous, isotropic FLRW and flat Minkowksi spacetimes. We therefore construct a modified invariant by multiplication with positive powers of the Weyl tensor, which vanishes on Weyl-flat spacetimes but remains singular on black-hole horizons. Because the power of derivatives in the numerator is higher than the denominator, the generic non-locality is milder. It remains to be understood what the effects on unitarity/stability and a well-defined initial-value problem are.

Our result 
 highlights two aspects. First,
our work highlights that the no-global-symmetries conjecture is based on assumptions that may not be generic in quantum gravity, because one can construct diffeomorphism-invariant actions which remove black-hole configurations from the path integral. Such actions might have a place in approaches to quantum gravity. The no-global-symmetries conjecture, being closely tied to the Einstein action and the Bekenstein-Hawking area law for the entropy, may thus be a result of taking the semi-classical approximation to a full quantum gravitational path integral too seriously and extrapolating it beyond its regime of validity. Specifically, the semi-classical approximation of black-hole evaporation predicts that a black hole evaporates fully without leaving behind a remnant. This may well be different if additional terms beyond the semi-classical Einstein term are considered in the action.  In particular, it is well-known that such terms can lead to regular black holes which possess an outer (event) horizon and an inner Cauchy horizon. During their Hawking evaporation process, such black holes reach a maximum temperature and subsequently cool down to leave behind a zero-temperature remnant asymptotically. It is currently not clear how a violation of the Bekenstein-Hawking area law and entropy bounds affects the conclusions of recent attempts to generalize the no-global-symmetries conjecture to broader frameworks for quantum gravity, beyond holography~\cite{Harlow:2020bee}.

Second,
our work highlights that removing black-hole configurations from the path integral comes at a high cost, because it requires dynamics which is non-local. Whether the non-localities even invalidate the theory (e.g., due to violations of causality and/or unitarity) remains to be investigated. We are of course not the first to propose that non-localities have an important role to play in gravity. Non-localities are well studied in phenomenological models for cosmology, because they may give rise to large effects in late-time cosmology, see, e.g.,~\cite{Deser:2007jk,Amendola:2017qge}, and they have been proposed to resolve the Big-Bang-singularity, e.g., in~\cite{Biswas:2010zk}. In many of these settings, stronger non-localities than in our theory are considered in the form of $\Box^{-1}$ and related operators.\\
In addition, non-local actions appear in the continuum limit of loop quantum gravity and spin foams~\cite{Borissova:2022clg,Borissova:2023yxs} and they have been investigated even in the context of asymptotic safety~\cite{Machado:2007ea}. There, non-localities may be difficult to reconcile with obtaining a predictive theory, because non-local terms with couplings of increasingly positive mass dimension likely give rise to an increasing number of Renormalization-Group (RG) relevant couplings. Our action is different in this respect, because the coupling multiplying $(C^2)^8/\chi^2$ has mass-dimension eight and is therefore RG irrelevant.

In this note, we have limited ourselves to removing black-hole configurations which are spherically symmetric from the path integral. However, there are of course other configurations of spacetime that are tied to violations of global symmetries. These are black-hole spacetimes with fewer Killing vectors, as well as wormhole configurations.

Going beyond spherically symmetric spacetimes, the mechanism to dynamically suppress black-hole configurations in the path integral can in principle be generalized. Horizon-detecting invariants for the Kerr spacetime exist, see, e.g.,~\cite{Abdelqader:2014vaa,Page:2015aia}; however $\chi$ is not a horizon detector. Accordingly, we conjecture that an action which suppresses both spherically symmetric black-hole spacetimes, as well as axisymmetric black-hole spacetimes, has to be constructed from (at least) two contributions. Further, horizon detectors for black-hole spacetimes without Killing vectors are, to the best of our knowledge, not known.
More generally, it is, to the best of our knowledge, also an open question whether a single curvature invariant exists, that detects the horizon of any black hole irrespective of the existence of any Killing vectors; a perusal of the relevant literature makes this appear unlikely, though. To achieve a complete suppression of black-hole configurations in the path integral, these questions must be answered in a first step. 

(Euclidean) wormholes also play an important role in the no-global-symmetries conjecture~\cite{Giddings:1987cg,Lee:1988ge,Abbott:1989jw,Coleman:1989zu,Kallosh:1995hi}, see~\cite{Hebecker:2018ofv} for a review. 
They may of course be removed from the path integral simply by declaring that the topology of spacetime is fixed. In settings, where this option is not attractive, one can invoke a mechanism similar to the one that we explored.
For instance, some wormhole configurations may be suppressed through higher-derivative terms in the action~\cite{Chojnacki:2021xtr}, whereas for others, similar horizon-detecting invariants to the ones we explored are relevant~\cite{McNutt:2021esy}. The interplay of topology change and dynamics beyond the Einstein-Hilbert action is, to the best of our knowledge, not well explored. However, the above examples highlight that topology change may potentially be suppressed dynamically by an appropriate choice of action.~\footnote{In the presence of matter, at least some topology changing configurations may be suppressed because the action for matter may diverge at the spacetime point where topology change occurs, see, e.g.,~\cite{Dowker:1997hj}.}

Let us note that it would be worthwhile to investigate the black-hole entropy in the presence of these non-local terms in the action. Given that the action diverges on the horizon, if Wald's entropy formula~\cite{Iyer:1994ys} applies, one might expect a diverging entropy; see~\cite{Platania:2023uda} for a discussion of non-local actions and resulting divergent black-hole entropy. This, in turn, is what one would expect for black holes that do not violate global symmetries, cf.~\cite{Harlow:2020bee}.  

Finally, let us comment on potential phenomenological and observational implications of the actions investigated in this note. These actions by construction would lead to the suppression of generic classes of black holes in the path integral, due to their sensitivity to apparent horizons of black holes. Accordingly, the observed physical spacetimes, which come with non-zero probability, are likely to be described by horizonless objects for which the action remains finite, but which are nevertheless sufficiently compact to reproduce qualitative features of classical black holes. Such black-hole mimickers have been suggested at various places in classical and quantum gravity, cf.~e.g.~\cite{Mazur:2001fv,Visser:2003ge,Mathur:2005zp,Almheiri:2012rt,Buoninfante:2024oxl, Eichhorn:2022bgu,Platania:2023srt,Holdom:2016nek,Carballo-Rubio:2023mvr, Arrechea:2023oax,Borissova:2022jqj,Liu:2024zti}. Potential observable signatures of spacetimes with horizon-like structures are gravitational-wave echoes~\cite{Cardoso:2016rao,Abedi:2016hgu,Barcelo:2017lnx,Cardoso:2017njb} and excess emission in the central brightness depression of a black-hole shadow~\cite{Vincent:2020dij,Eichhorn:2022fcl,Carballo-Rubio:2022aed}.

\begin{acknowledgments}

JB is supported by a doctoral scholarship by the German Academic Scholarship Foundation and an NSERC grant awarded to Bianca Dittrich. AE is supported by a Villum Grant under grant no 29405. SR is supported by the Deutsche Forschungsgemeinschaft (DFG) through the Walter Benjamin programme (Grant No. RA3854/1-1, Project ID No. 518075237). Research at Perimeter Institute is supported in part by the Government of Canada through the Department of Innovation, Science and Economic Development Canada and by the Province of Ontario through the Ministry of Colleges and Universities.

\end{acknowledgments}

\bibliographystyle{jhep}
\bibliography{references}

\end{document}